\documentclass[twocolumn,aps,prl,superscriptaddress,english]{revtex4-2}

\usepackage{graphicx}
\usepackage{dcolumn}
\usepackage{bm}
\usepackage{hyperref}
\usepackage{lipsum}
\usepackage{tikz,tikz-cd}
\usepackage{braket}
\usepackage{siunitx}
\usepackage{amsmath,amssymb,amsfonts,stmaryrd}

\hypersetup{
colorlinks=true,
linkcolor=blue,
filecolor=blue,
citecolor=blue,  
urlcolor=blue,
}

\renewcommand{\geq}{\geqslant}

\begin{document}

\title{Efficient fault-tolerant implementations of non-Clifford gates \\with reconfigurable atom arrays}

\author{Yi-Fei Wang}
\affiliation{Institute for Advanced Study, Tsinghua University, Beijing 100084, China}
\author{Yixu Wang}
\affiliation{Institute for Advanced Study, Tsinghua University, Beijing 100084, China}
\author{Yu-An Chen}
\affiliation{International Center for Quantum Materials, School of Physics, Peking University, Beijing 100871, China}
\author{Wenjun Zhang}
\affiliation{Department of Physics and State Key Laboratory of Low Dimensional Quantum Physics, Tsinghua University, Beijing 100084, China}
\author{Tao Zhang}
\affiliation{Department of Physics and State Key Laboratory of Low Dimensional Quantum Physics, Tsinghua University, Beijing 100084, China}
\author{Jiazhong Hu}
\affiliation{Department of Physics and State Key Laboratory of Low Dimensional Quantum Physics, Tsinghua University, Beijing 100084, China}
\affiliation{Frontier Science Center for Quantum Information and Collaborative Innovation Center of Quantum Matter, Beijing, 100084, China}
\author{Wenlan Chen}
\affiliation{Department of Physics and State Key Laboratory of Low Dimensional Quantum Physics, Tsinghua University, Beijing 100084, China}
\affiliation{Frontier Science Center for Quantum Information and Collaborative Innovation Center of Quantum Matter, Beijing, 100084, China}
\author{Yingfei Gu}
 \email{guyingfei@tsinghua.edu.cn}
\affiliation{Institute for Advanced Study, Tsinghua University, Beijing 100084, China}
\author{Zi-Wen Liu}
 \email{zwliu0@mail.tsinghua.edu.cn}
\affiliation{Yau Mathematical Sciences Center, Tsinghua University, Beijing 100084, China}

\date{\today}

\begin{abstract}
To achieve scalable universal quantum computing, we need to implement a universal set of logical gates fault-tolerantly, for which the main difficulty lies with non-Clifford gates.
We demonstrate that several characteristic features of the reconfigurable atom array platform are inherently well-suited for addressing this key challenge, potentially leading to significant advantages in fidelity and efficiency. Specifically, we consider a series of different strategies including magic state distillation, concatenated code array, and fault-tolerant logical multi-controlled-$Z$ gates, leveraging key platform features such as non-local connectivity, parallel gate action, collective mobility, and native multi-controlled-$Z$ gates. Our analysis provides valuable insights into the efficient experimental realization of logical gates, serving as a guide for the full-cycle demonstration of fault-tolerant quantum computation with reconfigurable atom arrays. 
\end{abstract}

\maketitle


\section{Introduction}

The implementation of reliable large-scale quantum computing holds great promise for significant technological advancements but poses substantial challenges in practice, as quantum systems are inherently susceptible to noise and errors. A crucial idea for tackling this problem is quantum error correction (QEC) \cite{nielsen2000quantum,shor95,gottesman1997}, wherein the central element is quantum codes that encode the logical information of quantum systems. Logical error rates can be suppressed by the error detection and correction procedure. To implement large-scale general-purpose quantum computation in practice, we further need to be able to execute a universal set of quantum gates at the level of logical qubits fault-tolerantly.  
The most straightforward fault-tolerant logical gates are those implemented by transversal gates upon codes, which take the form of tensor products of gates acting on disjoint physical subsystems like individual code qubits. Unfortunately, a no-go theorem of Eastin and Knill  \cite{eastin2009restrictions} states
that transversal operators on any nontrivial QEC code
cannot be universal, which calls for other approaches for fault-tolerant (FT) logical gates.
In general, Clifford gates represent the ``easy'' part---they can be classically simulated efficiently \cite{gottesman1998heisenberg,nielsen2000quantum} and are relatively straightforward to protect and implement fault-tolerantly. However, to achieve universal quantum computation, it is necessary to include non-Clifford gates such as $T$ and CC$Z$ gates, which represent the main bottleneck. 
To address this problem, multiple frameworks have been proposed and developed, including magic state distillation (MSD) \cite{bravyi2005universal,bravyi2012magic,haah2017magic,Hastings2018distillation}, code concatenation \cite{jochym2014using,yoder2016universal}, and code switching \cite{anderson2014fault,butt2023fault}.

\begin{table*}[t]
  \centering
  \begin{tabular}{| c | c | c| c | c|} 
  \hline
  & Non-local connectivity & Parallel gate action & Collective mobility & Native multi-controlled-$Z$ \\ 
  \hline
  Magic state distillation & \checkmark & \checkmark &  \checkmark & \\ 
  \hline
  Concatenated code array & \checkmark & \checkmark & \checkmark & \\ 
  \hline
  FT multi-controlled-$Z$ codes& \checkmark & \checkmark  & \checkmark & \checkmark\\
  \hline 
\end{tabular}
  \caption{Summary of the major schemes for the efficient fault-tolerant implementation of non-Clifford gates considered in this article and the characteristic features of the reconfigurable atom array platform that can significantly enhance their efficiency. 
  The rows correspond to different schemes for fault-tolerant non-Clifford gates, and the columns correspond to features of the reconfigurable atom array platform.
  Here, 
  non-local connectivity refers to the reconfigurable architecture that allows non-local gates \cite{Bluvstein_2022}; 
  parallel gate action refers to the parallel grid illumination that realizes parallel single qubit rotations~\cite{Saffman2022, Bluvstein_2023}; collective mobility refers to the transport of multiple qubits via moving 2D acousto-optic deflectors (AOD), which can be used to perform parallel entangling C$Z$ gates in a zone with global Rydberg excitation laser \cite{Levine2019};
  native multi-controlled-$Z$ refers to the experimental realization of a multi-qubit gate by moving multiple atoms into Rydberg blockage regime, e.g. CC$Z$ by preparing three atoms in the nearest-neighbor blockade regime \cite{Levine2019}. 
  }
  \label{tab:1} 
\end{table*}

From a practical viewpoint, FT implementation of non-Clifford logical gates faces fundamental obstacles when the system architecture or interaction structure is restricted to two spatial dimensions (2D) or lower, which is more feasible in various experimental platforms.  In particular, 
it is well known that for 2D stabilizer codes \cite{BravyiKonig13} (such as the surface code \cite{bravyi1998quantum,Dennis_2002,fowler2012surface} which has been a leading candidate for realizing fault tolerance) 
and even subsystem codes \cite{PastawskiYoshida15}, gates that can be implemented transversally or indeed with constant-depth quantum circuits are restricted to the Clifford group. 
As a result, the implementation of non-Clifford gates, which are required for universality, is expected to be difficult with 2D locality due to the necessity of long-range interactions.
Here, we consider the reconfigurable atom array quantum processor \cite{Bluvstein_2023}, an emerging hardware architecture \cite{science.aah3778,science.aah3752} that enables highly parallel and dynamically all-to-all gates, thereby overcoming the aforementioned geometric locality constraint.

Specifically, we propose and analyze several hardware-efficient schemes for fault-tolerantly implementing non-Clifford gates with reconfigurable atom arrays. 
The primary ones that we will elaborate on include magic state distillation, concatenated code array, and FT logical multi-controlled-$Z$ gates. 
Remarkably, all of these approaches capitalize on certain characteristic features of the atom array experimental platform, particularly the reconfigurability and parallel efficient control, which enable significant advantages; see Table~\ref{tab:1} for a summary.
We will describe the implementation methods and analyze their experimental feasibility in detail, from which it will become evident how the native features of the platform are particularly favorable for implementing non-Clifford gates.

\section{Magic state distillation}

Magic state distillation (MSD) and injection constitutes a major approach to achieve FT universal logical gates.
Roughly speaking, the protocol refers to the procedure of distilling certain non-stabilizer states to arbitrary fidelity from noisy states (which may have suffered from storage error) offline, and directly ``injecting'' them  into the circuit to realize non-Clifford gates \cite{bravyi2005universal}, both steps using only Clifford gates.
This method is based on assuming ideal Clifford gates as their fault tolerance can be achieved straightforwardly, and focus on dealing with noisy non-Clifford resources. 



\begin{figure}
    \centering
    \includegraphics{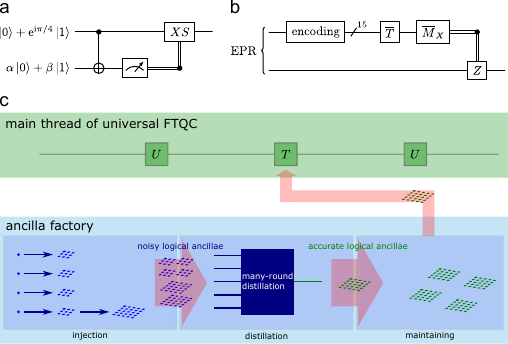}
    \caption{(a) The circuit for the implementation of a $T$ gate with an ancilla. Note that the input qubit is destructively measured and the ancillary qubit serves as the output. (b) Illustration of magic state distillation. One qubit of an entangled pair is encoded into the $\llbracket 15, 1, 3\rrbracket$ quantum Reed--Muller code and a logical $T$ gate is applied via transversal $T^\dag$ gates. Each $T^\dag$ gate is implemented using a noisy ancilla. After measurement on the 15-qubit code and a conditioned $Z$ on the other qubit of the EPR pair, the latter qubit is transformed to a more accurate ancilla. (c) Universal fault-tolerant quantum computation (FTQC) with magic state distillation. The ancilla factory supplies noisy ancillae that are injected to QEC codes of various sizes and undergo many-round distillation until the desired fidelity is achieved. The produced ancillae are then maintained by standard error correction procedure for quantum memory. When a logical $T$ is required in the main thread of the computation, a good ancilla is moved out from the factory to the computation region.} 
    \label{fig:Tdistillation}
\end{figure}

Here we consider the $T$ gate (i.e. $T= \exp(-\mathrm{i} \pi\sigma^z/8)$), a standard non-Clifford gate that forms a universal gate set together with Clifford gates. It can be implemented with the ancilla $\ket{T}=\ket{0}+\mathrm{e}^{\mathrm{i}\pi/4}\ket{1}$, as shown in Fig.~\ref{fig:Tdistillation}(a). Here, to distill the ancilla, we consider the scheme using the $\llbracket 15,1,3\rrbracket$ quantum Reed--Muller (QRM) code that has transversal logical $T$. 
We consider the distillation scheme shown in Fig.~\ref{fig:Tdistillation}(b) which consumes 15 noisy ancillae and outputs 1 more accurate ancilla. An EPR pair $(\ket{00}+\ket{11})/\sqrt{2}$ is prepared and one qubit is encoded into the 15-qubit code. Then a transversal $T$ gate is applied using the input noisy ancillae. Finally, all 15 qubits are measured in the $X$ basis. If any of the four $X$ stabilizers is not satisfied, the output will be discarded, otherwise one may apply a $Z$ operator conditioned on the product of all $X$ measurements which is exactly the logical $\bar{X}$ measurement \footnote{There is an alternative scheme: encode a $\ket{+}=(\ket{0}+\ket{1})/\sqrt{2}$ to the 15-qubit QRM code, apply a transversal $T$, and then decode to extract the information in the 15 qubits to one qubit. This scheme uses one less qubit but take twice the number of Clifford gates, which is slower and more sensitive to Clifford errors.}.  

Eventually, we would like to carry out magic state distillation on a logical level such that qubits in circuit \ref{fig:Tdistillation}(b) are protected by quantum codes, that is, all the ``qubit'' in the previous paragraph refers to logical qubit encoded in some codes (for example surface codes). The fault-tolerant universal quantum computation architecture using this logical level distillation is illustrated in Fig.~\ref{fig:Tdistillation}(c). A more feasible short-term goal is to distill $T$ ancillae on a physical level, as a demonstration of both the distillation scheme and the experiment techniques. 

For physical level distillation, since the $\llbracket 15,1,3\rrbracket$ QRM code is a 3D code, it is inefficient to implement the encoding using local gates in 2D since we need many swap gates for long-range CNOT gates, which not only takes more time but also introduces more errors. The reconfigurability of atom arrays can provide significant advantages. For distillation at the logical level using the surface code, non-local logical CNOT gates between two surface codes are required. Even if lattice surgery techniques \cite{horsman2012surface} are used to implement logical CNOT gate locally between adjacent code blocks, the non-locality in the distillation circuit still requires the additional logical swaps, each using 3 CNOT gates by lattice surgery, making the overhead much larger.

\begin{figure}
    \centering
    \includegraphics[width=1\linewidth]{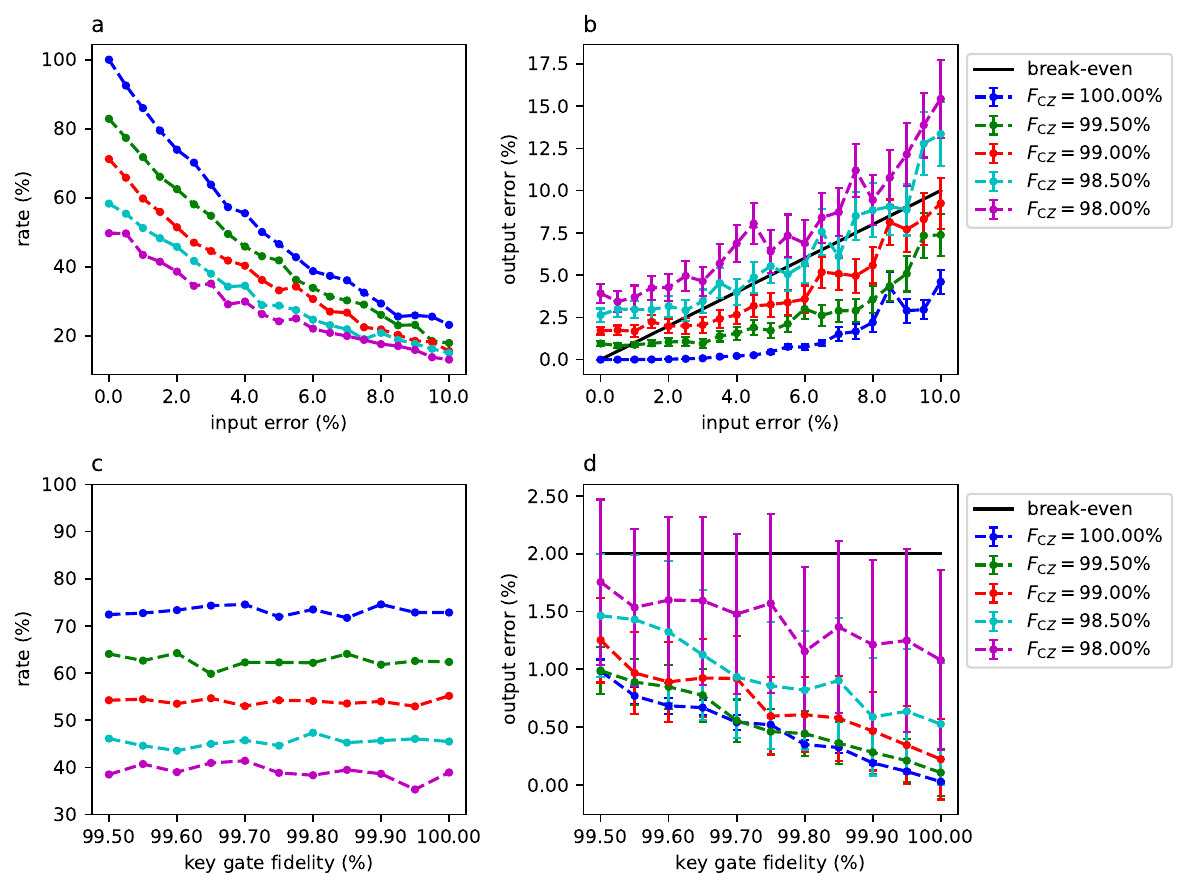}    
    \caption{Effect of two-qubit gate error on distillation of $T$ gates obtained from Monte Carlo simulation. (a, b) Rate of successful rounds and output noise as a function of input noise, under different fidelities (``$F_{\mathrm{C}Z}$'' in legend) of two-qubit C$Z$ gates with independent $Z$ errors on each qubit. Break-even condition that output noise equals to input noise is indicated by the black solid line in (b). (c, d) Rate of successful rounds and output noise as a function of key gate fidelity (see main text), under different fidelities of other C$Z$ gates. }
    \label{fig: distil numerical}
\end{figure}

To provide a first estimation of the feasibility of MSD on reconfigurable atomic systems, we consider a simplified error model where independent $Z$ errors can occur on each qubit when applying a C$Z$ gate. This simplification is based on the error analysis on realistic platforms \cite{Evered2023}. We simulate one distillation round 100 times at different input ancilla noise and C$Z$ gate fidelity. Fig.~\ref{fig: distil numerical}(a,b) reveal the performance of MSD when all C$Z$ gates have the same gate fidelity, which can serve as a reference for near-term experiments. Especially at the state-of-the-art C$Z$ gate fidelity $99.5\%$ \cite{Evered2023}, one can achieve break-even when the input infidelity is higher than $1\%$ ($\gtrsim 0.75\%$ according to our analytical result). 
Note that, since 1\% is much higher than the error of single-qubit rotation in recent techniques, distillation at the physical level serves more as a proof-of-principle demonstration than a practical procedure.

Fig.~\ref{fig: distil numerical}(c,d) reveals a remarkable observation: when the input noise is 2\%, a point at which 99\% C$Z$ fidelity achieves break-even (b), only by improving the fidelity of 5 key gates to 99.5\% can we achieve break-even when all other C$Z$ gates still have the fidelity of 98\%. In fact, our analytic computation shows that the linear dependence of the output error on the C$Z$ error comes totally from the 5 key gates. If these key gates have gate fidelity $(1-q)^2$, (that is, a $Z$ error can occur on each qubit with probability $q$ when applying the gate), while other C$Z$ gates have fidelity $(1-p)^2$, the leading order of the output error is $3.5q$. This linear dependence can be further suppressed using a flag protocol \cite{chao2018quantum,chao2018fault}; see Appendix A. 
Our analysis suggests that, at the fault-tolerant level, costs can be reduced by focusing on the improvement of these key gates, comparing with the former cost analysis where all C$Z$ gates are equally protected.

\section{Concatenated code array}

\begin{figure}
    \centering
    \includegraphics[scale=1.2]{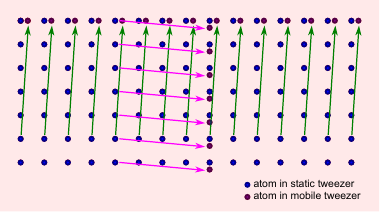}
    \caption{Parallel implementation of a universal set of logical gates by code concatenation. A logical qubit is encoded in the $\llbracket 7,1,3\rrbracket$ Steane code concatenated with the $\llbracket 15,1,3 \rrbracket$ QRM code forming a $7\times 15$ atom array (dark blue circles) held by an array of static tweezers generated with a spatial light modulator (SLM, red circles), where each row is a 15-qubit code and a physical ``qubit'' for the 7-qubit code. To apply CNOT gates transversally between two rows, say rows 1 and 6, one can use an array of mobile tweezers generated with AOD to shift one row of atoms to the neighbor sites of the other row (green arrow). One then turns off the tweezers, applies a row of Hadamard gates on the target line, turns on a pulse to apply C$Z$ gates on the atom pairs closer than the Rydberg blockade radius simultaneously, and applies Hadamard gates on the target line again. To apply physical CNOT gates simultaneously between two columns, say columns 5 and 9, one can move one column of atoms to the neighbor of the other (purple arrow). The other operations are similar.}
    \label{fig: concatenation array}
\end{figure}

Code concatenation offers another approach to bypass the Eastin-Knill theorem to achieve universal FT gates, which is also particularly fit for the atom array platform. The essential idea  is to ``combine'' different FT gate sets of different codes \cite{jochym2014using}. 
Consider two codes $\mathcal{C}_1$ and $\mathcal{C}_2$ such that the union of their transversal gate sets is universal. We concatenate these two codes such that each physical qubit in $\mathcal{C}_1$ is encoded as a logical qubit for  $\mathcal{C}_2$. For a small example, we can take $\mathcal{C}_1$ to be the $\llbracket 7,1,3\rrbracket$ Steane code and $\mathcal{C}_2$ the $\llbracket 15,1,3\rrbracket$ QRM code \cite{jochym2014using}.  $\mathcal{C}_1$ has transversal gate set $\{H, S, \mathrm{CNOT}\}$ while $\mathcal{C}_2$ has transversal gate set $\{T,\mathrm{CNOT}\}$. We can arrange the physical qubits into a $7\times 15$ array,  with each row forming the 15-qubit code while the collection of rows corresponding to the 7-qubit code; see Fig.~\ref{fig: concatenation array}. To implement a logical $S$ or CNOT, we can apply the gate qubit-wise: a qubit-wise $S$ is a logical $S^\dag$ for the 15-qubit code, and a qubit-wise $S^\dag$ is a logical $S$ for the 7-qubit code; similar is the CNOT gate. To implement a logical $T$, which is not transversal for the 7-qubit code, we need to apply 4 CNOT gates and 1 $T$ gate at the physical level of the 7-qubit code, which are transversal for the 15-qubit code: errors can only propagate within individual columns. To implement a logical $H$, which is transversal for the 7-qubit code but not transversal for the 15-qubit code, we need to apply a logical $H$ gate, which amounts to 14 CNOT gates and 1 $H$ gate, for each 15-qubit code: errors can only propagate within individual rows \cite{chamberland2017overhead}. Both $\mathcal{C}_1$ and $\mathcal{C}_2$ have distance 3 but neither has a transversal universal gate set. Nevertheless, we can implement a universal gate set fault-tolerantly in the concatenated code with an effective distance of 3.

The parallel gate action and the collective mobility features of the atom arrays are ideal for implementing a concatenated code array scheme. For instance, in the logical $T$ and $H$ implementation, 
CNOT gates between two rows or columns can be performed in parallel via transport-based entangling gates \cite{Bluvstein_2022}, see Fig.~\ref{fig: concatenation array} for details. To demonstrate the experimental feasibility, we give an estimation for the time cost of logical $T$ and $H$ based on the architecture and technology demonstrated in \cite{Bluvstein_2022}, utilizing a system of acoustic optical deflectors (AOD). This system enables a simultaneously movement of an entire row or column of the tweezers array. 
In the atom arrays, two atoms are separated by roughly \SI{10}{\micro\meter}. Two adjacent sites in static tweezer are separated by less than \SI{2}{\micro\meter}. 
The moving speed of atoms at which the fidelity is well preserved is roughly \SI{0.5}{m/s}. The typical moving time is then at the order of some \SI{20}{\micro\second}. Besides moving at a constant velocity there are other processes including the acceleration which has minor influence on time cost, the pulse implementing C$Z$ gates which lasts for roughly $\SI{200}{ns}\ll \SI{20}{\micro\second}$, and transferring between spatial light modulator (SLM) and AOD tweezers which takes roughly $100\sim \SI{200}{\micro\second}$ \cite{Bluvstein_2023}. Only the last procedure is relevant to our time estimation. For logical $T$ gate, 4 cycles of CNOT gates are needed \cite{jochym2014using, chamberland2017overhead}, involving row movements $R_7(7\rightarrow 6), R_6(6\rightarrow 1), R_6(1\rightarrow 6), R_7(6\rightarrow 7)$, where $R_i(j\rightarrow k)$ means moving the $i$th row of atoms from row $j$ to row $k$, taking roughly \SI{20}{\micro\second}, \SI{100}{\micro\second}, \SI{100}{\micro\second}, \SI{20}{\micro\second}, respectively, adding up to \SI{0.24}{ms} for moving only and \SI{0.84}{ms} with transferring time included (taking transferring time as \SI{150}{\micro\second}). For logical $H$ gate, 8 cycles of CNOT gates are needed \cite{chamberland2017overhead}. In the worst case that after each step columns are moved back to its original position, it takes roughly \SI{3.76}{ms} to implement the logical $H$ gate, comparing to an order of seconds for the decoherence time of an atom qubit. 
The time cost can be further reduced by optimizing the moving strategy based on different computational task at a software level, as well as using time optimal control techniques at a hardware level. 

\section{Fault-tolerant logical multi-controlled-$Z$ gates}

One advantage of the reconfigurable atom array platform is the natural physical implementation of multi-controlled-$Z$ gates, denoted by $\mathrm{C}^m Z$ where $m$ is the number of control qubits, which are non-Clifford when $m\geq 2$. Due to this feature, we are tempted to consider $\mathrm{C}^m Z$ gates which are suited to certain important scenarios (e.g.,~generating hypergraph states \cite{Rossi13:hypergraph} which are representative many-body entangled magic states \cite{LW22}) and generally provide an alternative choice of non-Clifford gates for circuit compilation.

Stabilizer codes based on triorthogonal matrices, such as the $\llbracket15, 1, 3\rrbracket$, $\llbracket49, 1, 5\rrbracket$, and a family of $\llbracket3k + 8, k, 2\rrbracket$ triorthogonal codes, support logical CC$Z$ gates implemented by transversal physical CC$Z$ gates \cite{bravyi2012magic, paetznick2013universal}. Additionally, the 3D surface code on the rectified cubic lattice, which exhibits a similar triorthogonal structure, has logical CC$Z$ gates implemented by transversal physical CC$Z$ gates \cite{Vasmer2019Three-dimensional}. This concept has been further generalized to the 4D octaplex tessellation, enabling the logical CCC$Z$ gate to be implemented by transversal physical CCC$Z$ gates \cite{JochymOConnor2020Four-dimensional}. Generally, the $D$-dimensional toric code permits logical non-Pauli gates from the $D$-th level of the Clifford hierarchy \cite{BravyiKonig13}. The duality between color codes and toric codes \cite{kubica2015unfolding} enables logical $\mathrm{C}^{D-1} Z$ gates in the $D$-dimensional toric code through transversal $R_m$ gates up to a Clifford circuit, where $R_D :=\text{diag}(1, \exp({{2\pi \mathrm{i}}/{2^D}}))$, saturating the Bravyi--K\"{o}nig bound \cite{BravyiKonig13}. 
Furthermore, we consider the  $D$-dimensional $(1,D-1)$-toric code on the hypercubic tessellation where the physical system consists of one qubit per edge, and the stabilizers are $X$-star (product of $X$ incident at a vertex) and $Z$-plaquette (product of $Z$ around a face) terms. It contains $0$-dimensional excitations (i.e., particles) and $(D-2)$-dimensional excitations. 
As discussed in detail in Appendix~B,
the logical $\mathrm{C}^{D-1} Z$ gates can be implemented fault-tolerantly with a constant-depth circuit of physical $\mathrm{C}^{D-1} Z$ gates.
This approach has the advantage that the implementation is straightforward and can be generalized directly to higher dimensions, without the need for intricate higher-dimensional rectifications or tessellations. 
It is worth emphasizing the suitability of high-dimensional codes and multi-controlled-$Z$ gates for the reconfigurable atom array platform. 
To achieve universality, we may use such codes in code switching or code concatenation strategies.  In this platform, these exotic high-dimensional codes can offer unique implementation advantages and greater flexibility for gate choice, further enhancing their utility in practical quantum computing.


\section{Discussion and Outlook}

Implementation of non-Clifford gates is costly but indispensable for fault-tolerant universal quantum computation. In this article, we describe how the native features of the reconfigurable atom array platform can lead to unique advantages in fault-tolerantly implementing non-Clifford gates. In particular, we provide detailed analyses for magic state distillation and code concatenation methods. Moreover, motivated by the unique feasibility of multi-controlled-$Z$ gates in this platform, we specifically discuss codes that use them to realize FT logical multi-controlled-$Z$.

Besides the methods analyzed in detailed in this article, there are other schemes for FT universal gates. A well established one is code switching \cite{bombin2015gauge,anderson2014fault}, which enables transversal universal gates through gauge fixing. 
This approach also inevitably involves codes beyond 2D so the reconfigurability of the atom array is again crucial.  It could also be worthwhile to further explore the usage of relevant methods such as flag qubits \cite{chao2018quantum,chao2018fault} and just-in-time decoding \cite{bombin2018transversal,Brown_2020}.

On the other hand, it would be valuable to systematically benchmark and compare the resource costs of different approaches for fault tolerance in the reconfigurable atom platform, in light of the comparison between e.g.~MSD and code switching with color codes \cite{beverland2021cost} in the literature. 

There are numerous other proposals exploring different features for reconfigurable atom  array platform, including biased noise \cite{cong2022hardware,sahay2023high}, erasure error conversion \cite{wu2022erasure} and highly non-local quantum LDPC codes \cite{xu2023constantoverhead}. With the rapid advancements of experimental technologies, now is an opportune time to explore and implement different methods which may pave the way for practical quantum computing.

\section*{Acknowledgements}

We thank Hui Zhai for comments on the draft and collaborations on related projects. YFW and YG are supported by NSFC12342501. WZ, TZ, JH and WC are supported by National Key Research and Development Program of China (2021YFA0718303, 2021YFA1400904) and National Natural Science Foundation of China (92165203, 61975092, 11974202). YW
is supported by Shuimu Tsinghua Scholar Program of Tsinghua University.

\bibliography{rydberg-ft-qec}

%

\end{document}


\title{Supplemantal Material: \\Efficient fault-tolerant implementations of non-Clifford gates \\with reconfigurable atom arrays}

\author{Yi-Fei Wang}
\affiliation{Institute for Advanced Study, Tsinghua University, Beijing 100084, China}
\author{Yixu Wang}
\affiliation{Institute for Advanced Study, Tsinghua University, Beijing 100084, China}
\author{Yu-An Chen}
\affiliation{International Center for Quantum Materials, School of Physics, Peking University, Beijing 100871, China}
\author{Wenjun Zhang}
\affiliation{Department of Physics and State Key Laboratory of Low Dimensional Quantum Physics, Tsinghua University, Beijing 100084, China}
\author{Tao Zhang}
\affiliation{Department of Physics and State Key Laboratory of Low Dimensional Quantum Physics, Tsinghua University, Beijing 100084, China}
\author{Jiazhong Hu}
\affiliation{Department of Physics and State Key Laboratory of Low Dimensional Quantum Physics, Tsinghua University, Beijing 100084, China}
\affiliation{Frontier Science Center for Quantum Information and Collaborative Innovation Center of Quantum Matter, Beijing, 100084, China}
\author{Wenlan Chen}
\affiliation{Department of Physics and State Key Laboratory of Low Dimensional Quantum Physics, Tsinghua University, Beijing 100084, China}
\affiliation{Frontier Science Center for Quantum Information and Collaborative Innovation Center of Quantum Matter, Beijing, 100084, China}
\author{Yingfei Gu}
 \email{guyingfei@tsinghua.edu.cn}
\affiliation{Institute for Advanced Study, Tsinghua University, Beijing 100084, China}
\author{Zi-Wen Liu}
 \email{zwliu0@mail.tsinghua.edu.cn}
\affiliation{Yau Mathematical Sciences Center, Tsinghua University, Beijing 100084, China}

\date{\today}

\maketitle




\onecolumngrid

\appendix

\section{A. Clifford errors in magic state distillation}\label{sec: MSD error}

In this appendix we discuss in some detail the effect of Clifford errors in magic state distillation of $T$ ancilla. We use the $\llbracket 15,1,3\rrbracket$ quantum Reed--Muller code with check matrix 

\begin{equation}
    \begin{aligned}
        H_X &= \left[\begin{tabular}{ccccccccccccccc}
        1 & 0 & 1 & 0 & 1 & 0 & 1 & 0 & 1 & 0 & 1 & 0 & 1 & 0 & 1 \\
        0 & 1 & 1 & 0 & 0 & 1 & 1 & 0 & 0 & 1 & 1 & 0 & 0 & 1 & 1 \\
        0 & 0 & 0 & 1 & 1 & 1 & 1 & 0 & 0 & 0 & 0 & 1 & 1 & 1 & 1 \\
        0 & 0 & 0 & 0 & 0 & 0 & 0 & 1 & 1 & 1 & 1 & 1 & 1 & 1 & 1 
        \end{tabular}\right], \\ 
        H_Z &= \left[\begin{tabular}{ccccccccccccccc}
        1 & 0 & 1 & 0 & 1 & 0 & 1 & 0 & 1 & 0 & 1 & 0 & 1 & 0 & 1 \\
        0 & 1 & 1 & 0 & 0 & 1 & 1 & 0 & 0 & 1 & 1 & 0 & 0 & 1 & 1 \\
        0 & 0 & 0 & 1 & 1 & 1 & 1 & 0 & 0 & 0 & 0 & 1 & 1 & 1 & 1 \\
        0 & 0 & 0 & 0 & 0 & 0 & 0 & 1 & 1 & 1 & 1 & 1 & 1 & 1 & 1 \\
        0 & 0 & 1 & 0 & 0 & 0 & 1 & 0 & 0 & 0 & 1 & 0 & 0 & 0 & 1 \\
        0 & 0 & 0 & 0 & 1 & 0 & 1 & 0 & 0 & 0 & 0 & 0 & 1 & 0 & 1 \\
        0 & 0 & 0 & 0 & 0 & 0 & 0 & 0 & 1 & 0 & 1 & 0 & 1 & 0 & 1 \\ 
        0 & 0 & 0 & 0 & 0 & 1 & 1 & 0 & 0 & 0 & 0 & 0 & 0 & 1 & 1 \\ 
        0 & 0 & 0 & 0 & 0 & 0 & 0 & 0 & 0 & 1 & 1 & 0 & 0 & 1 & 1 \\
        0 & 0 & 0 & 0 & 0 & 0 & 0 & 0 & 0 & 0 & 0 & 1 & 1 & 1 & 1
        \end{tabular}\right],
    \end{aligned}
\end{equation}    
where each row in $H_X$ defines an $X$ stabilizer which has the identity operator $I$ on sites with 0 while $X$ on sites with 1. For example, the first row in $H_X$ gives the stabilizer $X_1X_3X_5X_7X_9X_{11}X_{13}X_{15}$. Similarly, rows in $H_Z$ define the $Z$ stabilizers. One logical qubit encoded in this code is 
\begin{equation}
\begin{aligned}
    \ket{\Bar{0}}= \prod_{i=1}^4 \frac{I+S_X^i}{\sqrt{2}} \ket{00\cdots 0},\quad
    \ket{\Bar{1}} = \prod_{i=1}^4 \frac{I+S_X^i}{\sqrt{2}} \ket{11\cdots 1},
\end{aligned}
\label{logical qubit in 15 RM code}
\end{equation}
where $S_X^i$ is the $i$th $X$ stabilizer. It is straight forward to verify that 
\begin{equation}
    T^{\dag\otimes 15}\ket{\Bar{0}} = \ket{\Bar{0}},\quad T^{\dag\otimes 15}\ket{\Bar{1}} = \mathrm{e}^{\mathrm{i}\pi/4}\ket{\Bar{1}},
\end{equation}
indicating that this code has a transversal logical $T$ implementation via a bit-wise physical $T^\dag$ gate.

\begin{figure}
    \centering
    \includegraphics[scale=0.8]{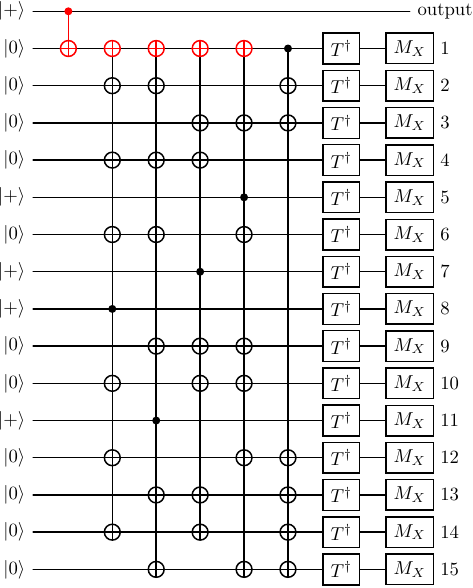}
    \caption{Circuit for magic state distillation of $T$ ancilla, adapted from \cite{fowler2012surface}. The key gates, CNOT between output and qubit 1, $\mathrm{CNOT}_{8,1}, \mathrm{CNOT}_{11,1},\mathrm{CNOT}_{7,1}, \mathrm{CNOT}_{5,1}$ are labelled red.}
    \label{fig: distillation circuit detail}
\end{figure}

The detailed circuit for distillation is shown in Fig.~\ref{fig: distillation circuit detail}. As discussed in the main text, we consider the major class of error, that is, the $Z$ error on the C$Z$ gates, which is modelled as 
\begin{equation}
    E(\rho) = \mathrm{C}Z \left((1-p)^2\rho +  p (I\otimes Z \rho I\otimes Z + Z\otimes I \rho Z\otimes I)+p^2 Z\otimes Z \rho Z\otimes Z\right) \mathrm{C}Z^\dag.
\end{equation}
The Choi gate fidelity is $F_{\mathrm{C}Z} = (1-p)^2$. A CNOT gate can be obtained from a C$Z$ gate by conjugating an $H$ on the target qubit, which converts the $Z$ error to an $X$ one. In this error model, we see that
\begin{itemize}
    \item The $Z$ error on the control qubit when entangling the output qubit with qubit 1 will directly come into the final result, yielding a $Z$ error.
    \item The five $X$ errors on the target qubit 1 will be spread to qubits 1, 2, 3, 12, 13, 14, 15 as $X_1X_2X_3X_{12}X_{13}X_{14}X_{15}$ since $\mathrm{CNOT}(XI)\mathrm{CNOT} = XX$. Since $T^\dag X = \mathrm{e}^{-\mathrm{i}\pi /4}XST^\dag$, where the factor is irrelevant while acting an $X$ before measuring $X$ has no effect, this error is equivalent to acting $S^{\otimes 7}$ on the 7 qubits. A straightforward calculation using equation (\ref{logical qubit in 15 RM code}) shows that this is a logical $S^\dag$ gate, which will be teleported to the output qubit. An $S^\dag$ error with probability $q$ contributes $0.5q$ to the output error.
    \item Other errors, including those in implementing $T^\dag$ using noisy ancillae, are not spread, reducing linearly the rate of success while the contribution to the output error is at a higher order.
\end{itemize}
From the first two points, we see that if the gate fidelity is $(1-p)^2$, the output qubit will be found with a $Z$ error at probability $p$, and an $S$ error at probability $5p$, yielding an output fidelity $1-3.5p$. Higher order contribution can come from two-qubit gates other than these 5 key gates.

\begin{figure}
    \centering
    \includegraphics[scale=0.8]{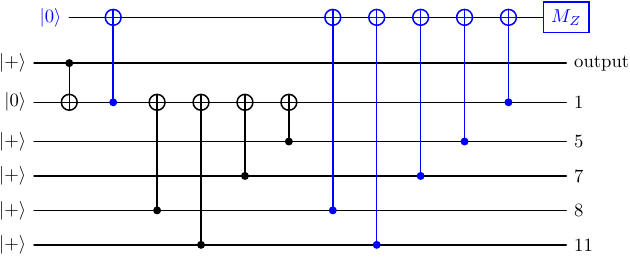}
    \caption{Using flag qubit gadgets to detect errors on qubit $1$. Here only the relevant gates in the distillation circuit are shown. The flag qubit gadgets are colored blue. If an $X$ error occurs on qubit $1$ within the flag qubit gadgets, their corresponding flag qubit will be measured $-1$ and this round of distillation should be discarded. However, a $Z$ error on the control qubit (1) of the leftmost blue CNOT gate can be propagated to qubits 1, 5, 7, 8, 11, which is a logical $Z$ operator. As a result, this flag gadget can reduce the linear dependence of the output error from $3.5p$ to $2.5p$.}
    \label{fig:flagcircuit}
\end{figure}    

\begin{figure}
    \centering
    \includegraphics[scale=.7]{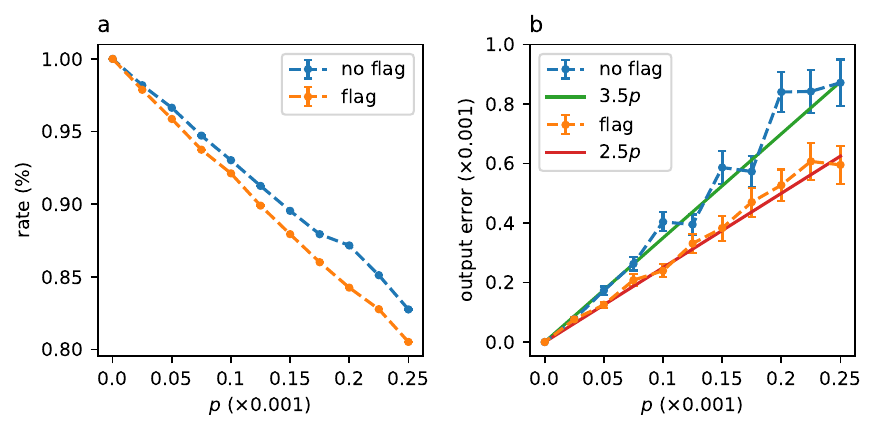}
    \caption{The rate (a) and output error (b) with and without flag gadgets, with data obtained by 200 rounds Monte Carlo simulation for each point. The input ancillae are accurate. The rate is suppressed linearly with flag since several single CNOT error now contributes linearly to the $-1$ flag measurement instead of the output error. The output error shows a behaviour of $3.5p$ and $2.5p$ with or without flag gadget, respectively.}
    \label{fig:Tflag}
\end{figure}   

We can use a flag gadget to further reduce this linear dependence, see the blue part of Fig.~\ref{fig:flagcircuit}. This flag gadget can detect whether there is an error on qubit 1 from CNOT gates between qubit 1 and qubits 5, 7, 8, 11, hence eliminate the contribution to the output error from these four gates. However, error on qubit 1 from the first CNOT between qubit 1 and the flag qubit can contribute linearly to the output error, since this error is propagated by the four following CNOT gates to a logical $Z$ error. Therefore, our flag gadget can reduce the number of key gates to 2 and reduce the output error from $3.5p$ to $2.5p$, where $(1-p)^2$ is the fidelity of the key gates. See Fig.~\ref{fig:Tflag} for a numerical simulation.

\section{B. Fault-tolerant logical $\mathrm{C}^{D-1} Z$ gates in $D$-dimensional toric codes}\label{sec: CnZ gates}

This section introduces a simple method for topologically protected FT logical $\mathrm{C}^{D-1} Z$ gates in $D$-dimensional toric codes using physical $\mathrm{C}^{D-1} Z$ gates. As an example, we start with two layers of 2D toric codes on the square lattice. One logical $\overline{X}_1$ gate in the first layer and another logical $\overline{Z}_2$ in the second layer are
\begin{equation}
    \begin{split}
        \overline{X}_1=\quad \vcenter{\hbox{\includegraphics[]{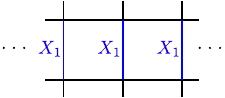}}}, \quad
        \overline{Z}_2= \quad \vcenter{\hbox{\includegraphics[]{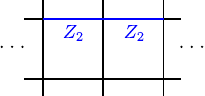}}}\quad.
    \end{split}
\end{equation}
The $\mathrm{C}Z$ gate between logical qubits in the two different layers of toric codes is
\begin{equation}
    \begin{aligned}
        \overline{\mathrm{C}Z}_{1,2}=& \quad \vcenter{\hbox{\includegraphics[]{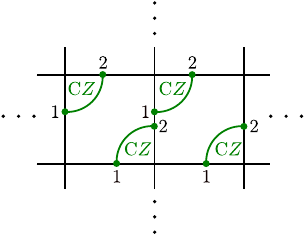}}}
        \quad,
    \end{aligned}
\end{equation}
the product of two physical $\mathrm{C}Z$ gates on each face, where the labels $1, 2$ indicate which layer it acts on. Two $\mathrm{C}Z$ gates correspond to two different paths from a corner of a square to the opposite corner.

This construction can be extended to three dimensions. Consider three layers of 3D toric codes. Logical $X$ gates become membrane operators, while logical $Z$ and $\mathrm{C}Z$ gates are the same as the 2D toric code.
Define logical $\overline{X}_3$ and logical $\overline{\mathrm{CC}Z}_{1,2,3}$ as
\begin{align}
        \overline{X}_3= \quad \vcenter{\hbox{\includegraphics{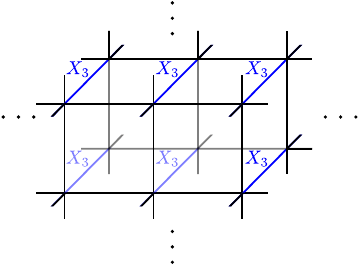}}} \quad, \\
        \nonumber\\ \overline{\mathrm{CC}Z}_{1,2,3}= \quad \vcenter{\hbox{\includegraphics[]{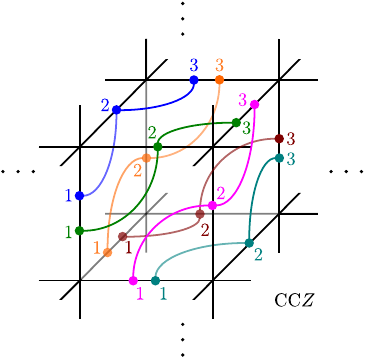}}}\quad.
\end{align}
where logical $\overline{\mathrm{CC}Z}_{1,2,3}$ is the product of six $\mathrm{CC}Z$ gates in each cube. The labels $1, 2, 3$ indicate which layer it acts on, and the six $\mathrm{CC}Z$ gates represent six paths from one corner to the opposite corner on a cube. 
One can verify that
\begin{equation}
    \overline{\mathrm{CC}Z}_{1,2,3} ~\overline{X}_3 ~\overline{\mathrm{CC}Z}_{1,2,3} = \overline{X}_3 ~\overline{\mathrm{C}Z}_{1,2}.
\end{equation}
This construction applies to the $D$-dimensional hypercube directly, where $\mathrm{C}^{D-1} Z$ gates act on the edges of each path from one vertex to the opposite corner.
In group cohomology language, the logical $\overline{\mathrm{C}Z}_{1,2}$ and $\overline{\mathrm{CC}Z}_{1,2,3}$ can be expressed by the cocycles $\frac{1}{2} a_1 \cup a_2 \in H^2(\mathbb{Z}_2 \times \mathbb{Z}_2, \mathbb{R} /\mathbb{Z})$ and $\frac{1}{2} a_1 \cup a_2 \cup a_3 \in H^3(\mathbb{Z}_2 \times \mathbb{Z}_2 \times \mathbb{Z}_2, \mathbb{R} /\mathbb{Z})$. In $D$ dimensions, the logical $\mathrm{C}^{D-1} Z$ gate corresponds to the cocycle $\frac{1}{2} a_1 \cup a_2 \cup \cdots \cup a_D \in H^D(\mathbb{Z}_2^D, \mathbb{R} /\mathbb{Z})$. The details can be found in Refs.~\cite{barkeshli2022higher, barkeshli2023codim,chen2023cup}.

\bibliography{rydberg-ft-qec}
